\documentclass{article}

\usepackage{arxiv}

\usepackage[utf8]{inputenc} 
\usepackage[T1]{fontenc}    
\usepackage{hyperref}       
\usepackage{url}            
\usepackage{booktabs}       
\usepackage{amsfonts}       
\usepackage{nicefrac}       
\usepackage{microtype}      
\usepackage{lipsum}
\usepackage{graphicx}
\usepackage{amsmath}
\usepackage{amssymb}
\usepackage{hyperref}

\usepackage{fdsymbol}

\newcommand{\medblackstardown}{%
\mathrel{\reflectbox{\rotatebox[origin=c]{180}{$\medblackstar$}}}}
\newcommand{\medwhitestardown}{%
\mathrel{\reflectbox{\rotatebox[origin=c]{180}{$\medwhitestar$}}}}

\usepackage[detect-none]{siunitx}
\graphicspath{ {./images/} }

\title{Simulation of Scanning Near-Field Optical Microscopy Spectra of 1D Plasmonic Graphene Junctions}

\author{
  Vyacheslav Semenenko\\
  Department of Electrical Engineering,\\
  University at Buffalo, Buffalo, NY 14260, USA\\
  \AND
  Mengkun Liu\\
  Department of Physics and Astronomy,\\
  Stony Brook University,\\
  Stony Brook, NY 11794, USA\\
  \AND
  Vasili Perebeinos\\
  Department of Electrical Engineering,\\
  University at Buffalo, Buffalo, NY 14260, USA\\
\texttt{vasilipe@buffalo.edu}\\
  
}

\begin{document}
\maketitle
\begin{abstract}
We present numerical simulations of scattering-type Scanning Near-Field Optical Microscopy (s-SNOM) of 1D plasmonic graphene junctions. A comprehensive analysis of simulated s-SNOM spectra is performed for three types of junctions. We find conditions when the conventional interpretation of the plasmon reflection coefficients from s-SNOM measurements does not apply. Our results are applicable to other conducting 2D materials and provide a comprehensive understanding of the s-SNOM techniques for probing local transport properties of 2D materials.
\end{abstract}


\section{Introduction}

Scattering type Scanning Near-Field Microscopy (s-SNOM) is a powerful
tool for studying local electronic properties of surfaces and 2D materials
in a few-nanometer-size domain~\cite{wang2016soliton,basov2019ldos,mengkun2018thz-nanoim}.
Simulation of the s-SNOM signal obtained in experiments is an essential
part of a technique for retrieving materials' physical properties.
The available mathematical modelings of s-SNOM experiments are
developed mainly for homogeneous surfaces~\cite{knoll2000basic,hillenbrand2001basic,cvitkovic2007elliptic-tip,basov2014tip-enh-spect,fogler2016gen-spec-method,mengkun2018scat-from-cone}. Similar calculations for the case of a composite wafer of planar
layers covered by graphene and exciting both by spherical~\cite{aizpurua2008substr-enh,basov2011ir-nanoscopy}  and cylindrical~\cite{mengkun2019snomthz}  tips have also been reported.
Besides,  one can find reports of simulated s-SNOM of 1D
graphene plasmon junctions~\cite{nikitin2013sic-step,nikitin2017corr}.

Currently, the most common application of s-SNOM is a visualization of electric properties of surfaces and atomically thin 2D materials. The s-SNOM images give a good qualitative understanding of a sensed object. However, retrieving accurate quantitative information on the length scale beyond the plasmon wavelengths is still a challenge. Some progress in this direction is offered by machine learning techniques~\cite{mengkun2021ml}. The plasmon's reflectance is one of the most valuable physical quantities of interest~\cite{nikitin2013sic-step,wang2016soliton}.
In graphene, plasmon reflectance was reported both analytically and numerically for various types of
1D junctions: discontinuously
changed doping in graphene supported by a homogeneous dielectric substrate~\cite{khavasi2015diff-cond,khavasi2016brewster},
homogeneously doped graphene sheet supported by a wafer with discontinuously
changed dielectric permittivity~\cite{bludov2018abrupt}, scattering
regions, such as a gap in graphene~\cite{nikitin2013foba,fogler2018junctions},
Gaussian-profile spatial distortion of graphene doping~\cite{nikitin2013foba}, and 1D corrugations of graphene sheet~\cite{nikitin2017corr}.

The electrical conductivity in 2D materials and the dielectric environment above and below define the plasmon wavelength and the electric field distribution—the difference in those physical parameters at two sides of the junction results in a finite reflection coefficient. Therefore, a reverse problem can be solved for the physical properties at one side of the junction using those at another side and reflection coefficient. Measuring the s-SNOM signal over a junction has certain advantages compared to getting the signal from unknown material and a known one used as a reference. The s-SNOM scan of the junction contains an immense amount of information, which is not apparent from simple fitting procedures. Therefore, a more significant amount of parameters can be retrieved by fitting them to a proper model function.

\section{Formulation of the Problem}

\begin{figure*}
\centering{\includegraphics[width=0.99\textwidth]{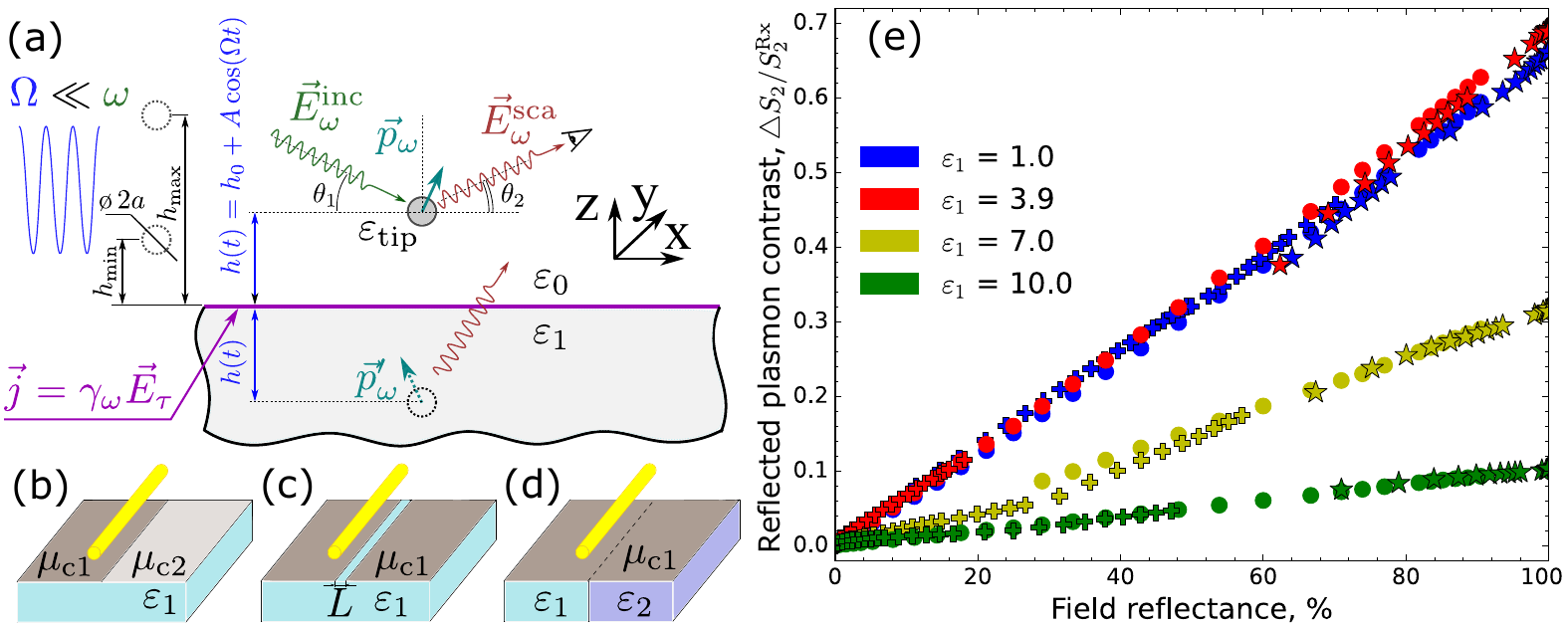}}

\caption{Schematics of s-SNOM of (a) flat dielectric covered by a homogeneous
conducting 2D material (see text for details),
and three plasmon junctions geometries considered in the paper: (b)
discontinuous doping, (c) gap in graphene, and (d) discontinuous wafer,
sensed by a cylinder s-SNOM tip (sketched in yellow). Panel (e) shows
filtered relations between plasmon reflection coefficient and simulated
measurements of the $S_2$ signal contrast (details are
explained in the text) for the three different
types of junctions and different dielectric permittivities $\varepsilon_{1}$
supporting graphene on the left side of the junctions and $\varepsilon_{2}$ on the right side in a case (d). The junction
types (b), (c), and (d) are marked with rounds, stars, and fat pluses,
correspondingly. }

\label{fig:s-snom-basics}
\end{figure*}

In this work, we develop a model for numerical simulation of the s-SNOM signal
of 2D conducting materials taking into account the signal's modulation from the oscillating cylindrical tip, as shown in Fig.~\ref{fig:s-snom-basics}a.
The conductivity
$\gamma_{\omega}$~\cite{xia2014mat,xia2015mat-recent} characterizes a 2D material. In the case of graphene, we assume the Drude model~\cite{hanson2008dyadic} for the conductivity.
Our simulations employ a model of s-SNOM of the bare plain surface of a
homogeneous dielectric following Refs.~\cite{knoll2000basic}
and \cite{hillenbrand2001basic}. That model was extended to enable
simulations of the s-SNOM of the layered planar structure of homogeneous
materials covered by graphene (or other 2D conducting material)~\cite{aizpurua2008substr-enh,basov2011ir-nanoscopy,mengkun2019snomthz}.
As in the latter references and unlike the original one Ref.~\cite{knoll2000basic}, we consider the cylindrical tip parallel to the
sample surface and thin compared to the characteristic size of the $E$-field inhomogeneity.

The solution in the Fourier domain makes the problem highly parallelizable. We have achieved an excellent  run-time simulation of the s-SNOM scans for three different types, as shown in Fig.~\ref{fig:s-snom-basics}b-d. We will refer to them as discontinuous conductivity, a gap in graphene, and a discontinuous wafer, correspondingly.
We simulated numerous scans for different parameters and verified a technique used by experimentalists for retrieving plasmon reflection coefficients~\cite{wang2016soliton}. The main goal of the present research is to determine under which conditions simple assumptions about the s-SNOM $S_2$ signal across a junction can retrieve a reflection coefficient of the plasmon. The latter can be calculated independently for those junctions~\cite{semenenko2020reflection} and compared with the $S_2$ signal contrast for different types of junctions and dielectric constants~$\varepsilon_{1}$
of the wafer supporting substrates, as shown in Fig.~\ref{fig:s-snom-basics}e.
While the results in Fig.~\ref{fig:s-snom-basics}e are more or less in agreement with the common-sense  assumption~\cite{wang2016soliton}, namely, the contrast of the $S_2$ signal is proportional to the reflection
coefficient; there are situations when those dependencies can be much more complex. Those cases are not shown in Fig.~\ref{fig:s-snom-basics}e, but are discussed below. The non-unique relationships between the s-SNOM signal contrast and reflection coefficient are explained by certain arbitrariness in how the s-SNOM scan of a junction can be processed.

\section{Basics of the s-SNOM simulations\label{sec:Basics-of-s-SNOM}}

In the s-SNOM setup, as shown in Fig.~\ref{fig:s-snom-basics}a,
an incident electromagnetic wave with the amplitude $\vec{E}_{\omega}^{\mathrm{inc}}$
scatters of the tip, which is closely located to the sensed surface.
The scattered radiation with the amplitude $\vec{E}_{\omega}^{\mathrm{sca}}$
(at the point where it is measured) is generated by the varying
tip's dipole moment and its reflection in the sensed sample. In the previous
models, the scattered far-field was calculated, and the tip's mechanical oscillations were taken into account
directly~\cite{knoll2000basic,hillenbrand2001basic,mengkun2018thz-nanoim}, where the angle between the sample's plane
and the incident TM polarized wave $\theta_{1}\approx0$ (see Fig.~\ref{fig:s-snom-basics}a), such that the dipole moment is directed
along the $z$-axis. In the case of a homogeneous surface, this assumption
is mathematically self-consistent, and the influence of the  mirrored dipole
$\vec{p}_{\omega}$ is reduced by a scaling factor
of the scattered amplitude $\vec{E}_{\omega}^{\textrm{sca}}$. Here, we extend those approaches  to the
non-homogeneous surface cases with plasmonic junctions (see Figs.~\ref{fig:s-snom-basics}b-d).
If the tip has polarizability in $x$-direction, the assumption that $\vec{p}_{\omega}\parallel z$-axis is no longer
valid even if $\theta_1\approx0^\circ$.
Therefore, our model, along with arbitrary $\theta_1$ and $\theta_2$, takes into account the tip's polarizability tensor and, correspondingly, an electric
field applied to the tip in $x$-direction. All the plots presented here are calculated for $\theta_1=0^\circ$ and $\theta_2=0^\circ$. The analysis of the angle dependence is beyond the scope of this work.

The components of the incident field amplitudes $\vec{E}_{\omega}^{\mathrm{inc}}$
are given by:
\begin{equation}
\vec{E}_{\omega}^{\mathrm{inc}}=\begin{bmatrix}E_{\omega,x}^{\mathrm{inc}}\\
E_{\omega,z}^{\mathrm{inc}}
\end{bmatrix}=E_{0}\begin{bmatrix}-\sin\theta_{1}\\
\cos\theta_{1}
\end{bmatrix},\label{eq:E-ref-dir}
\end{equation}
where $E_{0}$ is the incident wave's amplitude. For boosting the sensitivity and getting rid of the background signal, which mainly characterizes the tip's size and shape, the tip is forced to oscillate with a mechanical frequency $\Omega$ of the cantilever. Therefore, the s-SNOM
setup measures the time dependence of $\vec{E}_{\omega}^{\mathrm{sca}}\left[h\left(t\right)\right]$,
where $h\left(t\right)$ is the instant height of the tip's center
above the surface. The time-dependent signal is demodulated by a lock-in
amplifier producing the resulting signal's series of harmonics:
\begin{equation}
\begin{aligned}S_{n} & =\frac{2}{T}\int\limits _{0}^{T}\left(\vec{E}_{\omega}^{\mathrm{sca}},\vec{\tau}\right)\cos n\Omega t\,dt,\\
 & T=\frac{2\pi}{\Omega},\;\vec{\tau}=\begin{bmatrix}-\sin\theta_{2}\\
\cos\theta_{2}
\end{bmatrix},
\end{aligned}
\label{eq:Sn-def}
\end{equation}
where the round braces mean the scalar product, and $\vec{\tau}$
is a vector, which direction is determined by the angle $\theta_{2}$ (see Fig.~\ref{fig:s-snom-basics}a). As experimentalists ordinarily use the second-order
of the demodulated electric field, we show all our results for $n=2$.

\begin{figure}
\centering{\includegraphics[width=0.99\linewidth]{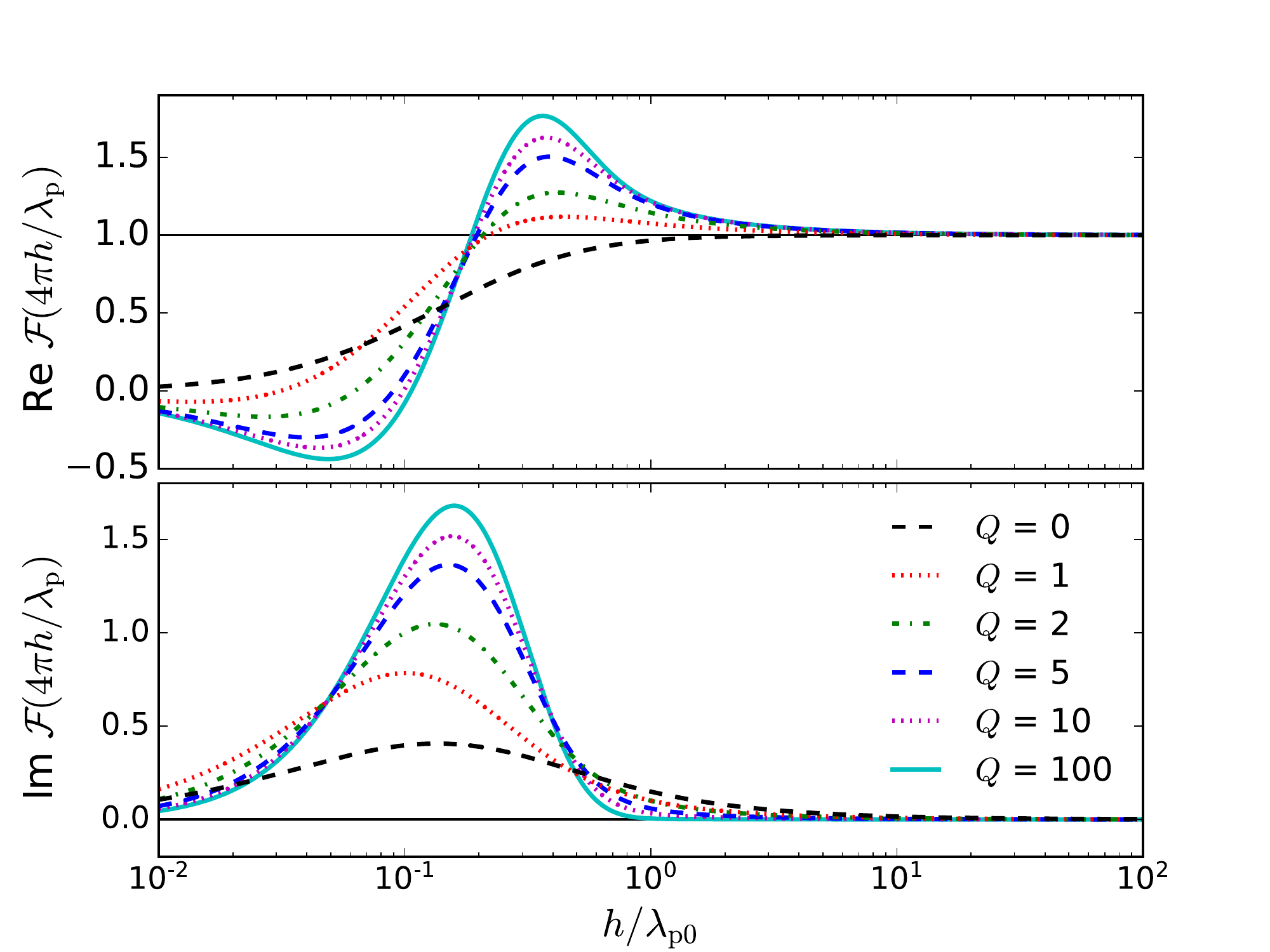}}

\caption{{\small{}$\mathcal{F}$ complex function versus normalized dipole
height $h$ above the sensed surface, calculated for different quality
factors of 2D charge density oscillations $Q$ (see text).}}

\label{fig:F-mathcal-scaling}
\end{figure}

The scalar product $\left(\vec{E}_{\omega}^{\mathrm{sca}},\vec{\tau}\right)$
in Eq.~(\ref{eq:Sn-def}) in the Fourier domain is proportional to
the amplitude $p_{\omega}$ of the dipole oscillations induced by
the incident reference wave and the interaction with the sensed object.
Therefore, the relationship between $\vec{p}_{\omega}$ and the input amplitude $\vec{E}_{\omega}^{\mathrm{inc}}$
of the incident wave determines the $S_{n}$ signal. Under the assumption of a thin tip, this relationship
can be derived if the tip's polarizability tensor $\hat{\alpha}$
and the spatial distribution of $E$-field $\vec{E}_{\omega}^{\mathrm{ind}}$ induced by the oscillating
dipole moment of the tip $\vec{p}_{\omega}$ are both known. The relationship between $\vec{E}_{\omega}^{\mathrm{ind}}$
at the center of the tip's dipole and the dipole moment $\vec{p}_{\omega}$ are given by:
\begin{equation}
\vec{E}_{\omega}^{\mathrm{ind}}=\frac{\hat{\beta}\left(h\right)\vec{p}_{\omega}}{2\varepsilon_{0}h^{2}},\label{eq:beta-tensor-def}
\end{equation}
where a tensor $\hat{\beta}$ depends on the properties of the sensed surface
and the height of the tip above it. In Ref.~\cite{knoll2000basic},
where s-SNOM of a bare homogeneous wafer was considered,
a similar expression for a point-dipole moment and the induced electric field
was derived for the case when $\vec{p}_{\omega}$ is normal to
the sensing plane. In that case, $\hat{\beta}$-tensor is reduced
to a scalar value
\begin{equation}
\beta=\frac{\varepsilon_{1}-\varepsilon_{0}}{\varepsilon_{1}+\varepsilon_{0}},\label{eq:beta-simple}
\end{equation}
where $\varepsilon_{0}$ and $\varepsilon_{1}$ are dielectric permittivities
above and below the conducting surface, correspondingly. It corresponds
to $\beta_{zz}$ component of the tensor in the case of sensing a homogeneous
bare semi-infinite wafer. Repeating the same derivation for $x$-components
of $\vec{p}_{\omega}$ and $\vec{E}_{\omega}^{\mathrm{ind}}$, one
can show that $\hat{\beta}=\beta\hat{I}$, where $\hat{I}$ is the unity
tensor. In our calculations, if not stated otherwise, we will consider $2\times2$ tensor's dimensions  due to the homogeneity in the $Y$-direction. Eq.~(\ref{eq:beta-tensor-def}) is solved together with the following relationship:
\begin{equation}
\vec{p}_{\omega}=\hat{\alpha}\left(\vec{E}_{\omega}^{\mathrm{inc}}+\vec{E}_{\omega}^{\mathrm{ind}}\right),
\end{equation}
where $\hat{\alpha}$ is the tip's polarizability tensor~\cite{novotny2003longitudinal}. In the case of a homogeneous,
perfectly conducting cylinder, the tensor $\hat{\alpha}$ is reduced to $\alpha\hat{I}$, where
$\alpha=\varepsilon_{0}a^{2}/2$, $a$ is tip's radius and $\hat{I}$ is the identity tensor.
%

Therefore, we obtain:
\begin{equation}
\vec{p}_{\omega}=\left(\hat{I}-\frac{\hat{\alpha}\hat{\beta}}{2\varepsilon_{0}h^{2}}\right)^{-1}\hat{\alpha}\vec{E}_{\omega}^{\mathrm{inc}},\label{eq:p-E-ref-rel-tensor}
\end{equation}
where $\left(...\right)^{-1}$ stands for matrix inversion. One can see
that the most challenging part of the calculations is finding
components of $\hat{\beta}$ tensor.

\begin{figure}[!h]
\centering{\includegraphics[width=0.99\columnwidth]{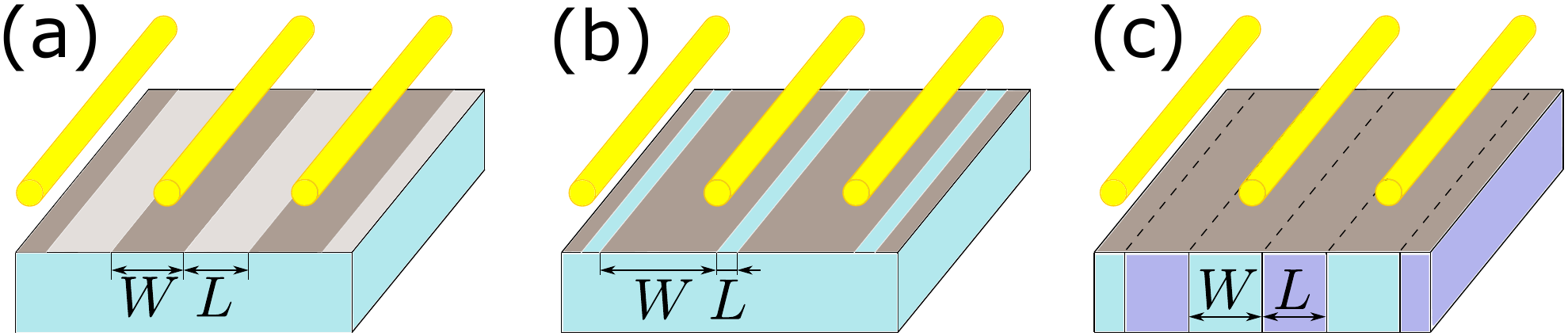}}

\caption{Periodic geometries (a-c) adapted for the numerical analysis we use
corresponding to the junctions depicted in Fig.~\ref{fig:s-snom-basics}b-d.}

\label{fig:periodic-geometry}
\end{figure}
The mirrored dipole moment $\vec{p}_{\omega}^{\prime}$ can be calculated as follows:
\begin{equation}
\vec{p}_{\omega}^{\prime}=\hat{\beta}^{\prime}\vec{p}_{\omega},\;\hat{\beta}^{\prime}=\beta\,\begin{pmatrix}-1 & 0\\
0 & 1
\end{pmatrix},\label{eq:beta-prime-tensor}
\end{equation}
where $\beta$ is a scalar factor given by Eq.~(\ref{eq:beta-simple}) and
expression for $\hat{\beta}^{\prime}$ applies for a homogeneous bare semi-infinite
wafer, which is the same for both small-ball- and thin-rod-like tips.
\begin{figure*}
\centering{\includegraphics[width=0.99\linewidth]{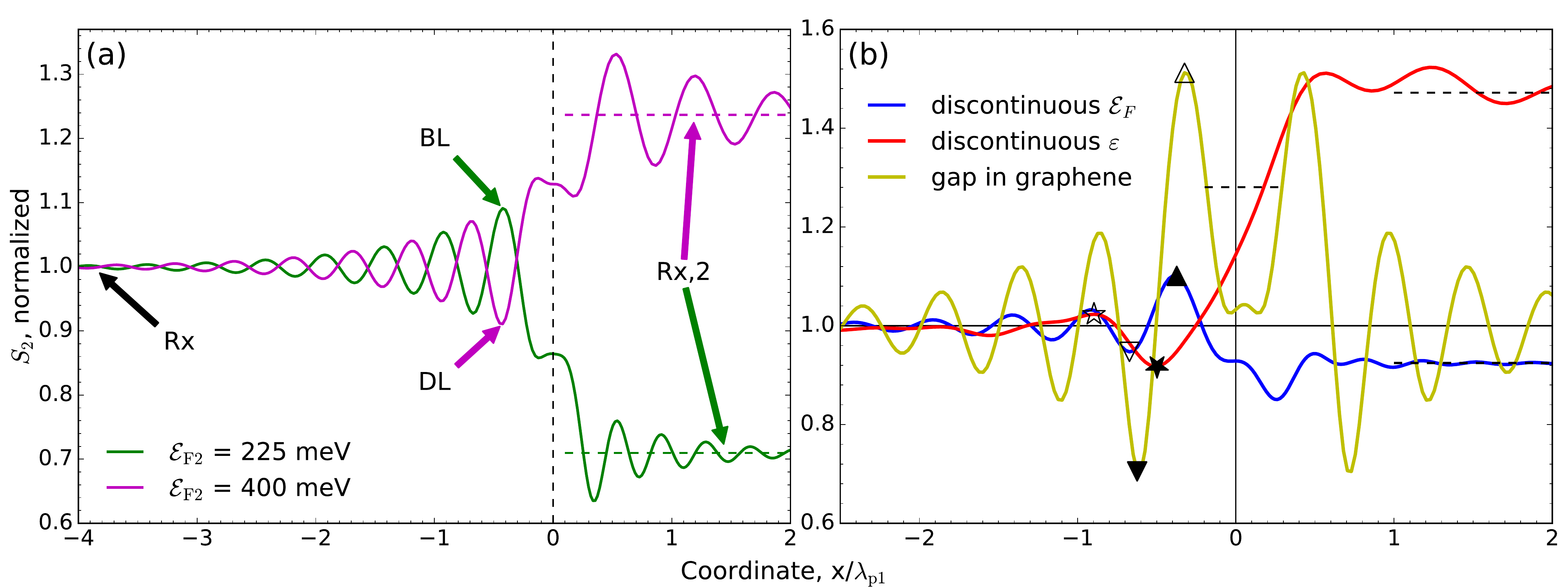}}
\caption{(a) Simulated $S_{2}$ scans calculated for two junctions of graphene
with discontinuous doping, giving the same reflection coefficient of the
plasmon. The relaxed signal value (Rx), brightest (BL), and darkest (DL)
lines positions are shown on the left side of the junction. Dielectric
permittivity of the wafer supporting the junctions is $\varepsilon_{1}=3.9$
and Fermi energy of graphene on the left side is $\mu_{\mathrm{c}1}=300$
meV in both cases. (b) Samples of $S_{2}$ scans of all three
types of junctions. Parameters of the junctions are the following:
(discontinuous doping) $\mu_{\mathrm{c}2}=210$ mV, (discontinuous
wafer) $\varepsilon_{2}=3.9$, (gap in graphene) $L=10$ nm. The simulations
are done at frequency $\omega=120$ meV, Fermi energy of the graphene
on the left side of a junction $\mu_{\mathrm{c}1}=300$ meV, the dielectric
permittivity of a wafer on the left side is $\varepsilon_{1}=7.0$,
electron scattering rate on both sides of a junction is the same
$\nu=10$ meV, minimum and maximum spans between the tip and the sample
are $h_{\mathrm{min}}-a=5$ nm, $h_{\mathrm{max}}-a=50$ nm, where $a=30$ nm is
the tip's radius. The whole
scans are normalized to the absolute value of the relaxed signal on
the junction's left side. The vertical dashed line shows the junction's
coordinate. Markers on panel (b) illustrate the classification rules
for brightest and darkest lines, which are used to calculate reflected
plasmon strength  following  Eq.~(\ref{eq:abs-dS2-def}). Triangle
and its variations are used when the plasmon wavelength on
the right side of a junction is less than one on the left side. Star
and its variations are used in the opposite case when $\lambda_{\mathrm{p}2}>\lambda_{\mathrm{p}1}$.
In the case of a gap in graphene junction, $\lambda_{\mathrm{p}2}=0$
is assumed. Straight oriented markers ($\medblacktriangleup$, $\medtriangleup$,
$\medblackstar$ and $\medwhitestar$) are plot for the case when
Eq.~(\ref{eq:abs-dS2-def}) is calculated using $S_{2}^{\mathrm{BL}}$.
Their overturned versions ($\medblacktriangledown$, $\medtriangledown$,
$\medblackstardown$, and $\medwhitestardown$
)  are plot when $S_{2}^{\mathrm{DL}}$
is used. A solid version of a marker designates a line for which a condition
in Eq.~(\ref{eq:abs-dS2-def}) is true, and the rest line then is designated with a marker of the same type, but it is opened and oppositely oriented version.}
\label{fig:S2-samples}
\end{figure*}
%

%
\begin{figure*}
\centering{\includegraphics[width=0.99\linewidth]{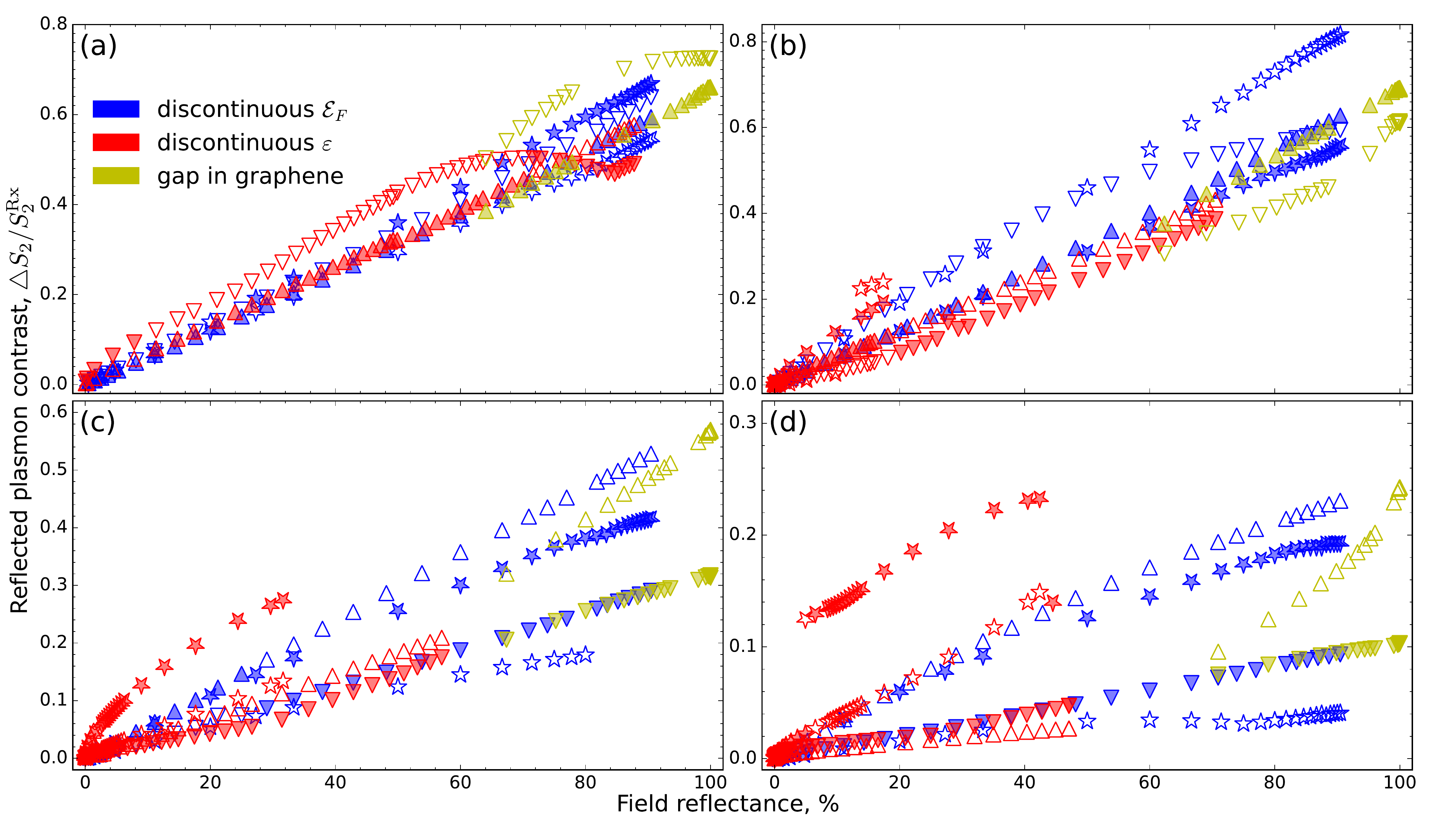}}

\caption{{\small{}$\left|S_{2}\right|$ signal contrast on
the left side of a junction, i.e. $S_2$ maximum normalized to its relaxed value $S_{2}^{\mathrm{Rx}}$,
versus modulus of the plasmon reflection coefficient for  different
dielectric permittivities of the medium below graphene. On the left
side of a junction: (a) $\varepsilon_{1}=1.0$, (b) $3.9$, (c) $7.0$ and (d) $10.0$.
These simulations are done at  fixed $h_{\mathrm{min}}=35$
nm and $h_{\mathrm{max}}=80$ nm. All the simulations use graphene's Fermi energy $\mu_{\mathrm{c}}=300$ meV (or $\mu_{\mathrm{c}1}=300$
meV on the left side of the junction for the case of discontinuous
doping junction), electron scattering rate in it $\nu=10$ meV, and
the tip's radius $a=30$ nm. The markers here have the same meanings as in Fig.~\ref{fig:S2-samples}.}}

\label{fig:S2-refl-scaling}
\end{figure*}

Before discussing plasmonic junctions, we consider a homogeneous sample to demonstrate $\hat{\beta}^{\prime}$ calculations.
Our analysis shows (see Appendix~\ref{sec:app-A}) that in the case of a homogeneous conducting surface covering
a homogeneous semi-infinite wafer, $\hat{\beta}$ is proportional
to $\hat{I}$ as in the case of a bare wafer (see Eq.~(\ref{eq:beta-simple})). However,
the coefficient $\beta$ is different, and it depends on the height $h$
of the dipole above the plane:
\begin{align}
 & \beta=\left[1-\frac{2\varepsilon_{0}}{\varepsilon_{0}+\varepsilon_{1}}\mathcal{F}\left(2q_{\mathrm{p}}h\right)\right],\;\label{eq:beta-analyt}\\
 & \mathcal{F}\left(\zeta\right)=\zeta^{2}e^{-\zeta}\left[\mathrm{Ei}\left(\zeta\right)+\pi i\right]-\zeta,\; \\
 & \mathrm{Ei}\left(\zeta\right)=\int\limits _{-\infty}^{\zeta}\frac{e^{u}}{u}du,\; q_{\mathrm{p}}=\frac{\left(\varepsilon_{0}+\varepsilon_{1}\right)\omega}{4\pi i\gamma_{\omega}}
 \label{eq:F-mathcal-scaling-main}
\end{align}
where $\gamma_{\omega}$ is complex conductivity of
the 2D surface, and $q_{\mathrm{p}}$ is the complex wavelength of the
2D charge density oscillations (or plasmons) supported by the surface
at frequency $\omega$. The complex function $\mathcal{F}\left(\zeta\right)$ in Eq.~(\ref{eq:beta-analyt})
has the following asymptotics:
$\mathcal{F}\left(0\right)\rightarrow0$ and $\mathcal{F}\left(\infty\right)\rightarrow1$.
In both limits, Eq.~(\ref{eq:beta-analyt}) reduces to Eq.~(\ref{eq:beta-simple}). In the first limit, one should substitute
$\varepsilon_{1}\rightarrow\infty$.
It means that at $h\ll\left|q_{\mathrm{p}}^{-1}\right|$ the dipole
interacts with the sample as if it was a perfectly conducting plane and
at $h\gg\left|q_{\mathrm{p}}^{-1}\right|$ as it was a bare dielectric
without conducting surface on the top. To show what happens for intermediate values of $\zeta$, in Fig.~\ref{fig:F-mathcal-scaling}  we show
the dependence of $\mathcal{F}$ on $h/\lambda_{\mathrm{p}0}$ for
different ``quality factors'' $Q$ of the plasmons excitation defined below.
Here we assume the Drude-like conductivity $\gamma_\omega$ of the 2D material that in the case of graphene is given by~\cite{hanson2008dyadic}:
\begin{equation}
\gamma_{\omega}=\frac{e^{2}\mathcal{E}_{\mathrm{F}}}{\pi\hbar^{2}\left(i\omega+\nu\right)},\label{eq:gcond-Drude}
\end{equation}
where $\mathcal{E}_{\mathrm{F}}$ is the Fermi energy and $\nu$ is the
electron scattering rate. The values of the quality factors $Q=\omega/\nu$ are given in the legend of Fig.~\ref{fig:F-mathcal-scaling} for all the cases except for the pure dissipative admittance, in which case $i\omega$ is set to zero in
Eq.~(\ref{eq:gcond-Drude}), i.e. $Q=0$. We define a characteristic
length $\lambda_{\mathrm{p}0}$ as: $\lambda_{\mathrm{p}0}=2\pi/\mathrm{Re}\,q_{\mathrm{p}}$
for $Q>0$, and $\lambda_{\mathrm{p}0}=-2\pi/\mathrm{Im}\,q_{\mathrm{p}}$, for $Q=0$.

Eq.~(\ref{eq:beta-analyt}) corresponds to the case of the reflected near-field signal at the $\vec{p}_{\omega}$ dipole position.
If one needs to calculate the field above the tip, at some height
$H$ above the sample, the same $\mathcal{F}$-function can be used with $2h$ replaced by
$h+H$ (see the derivation of
Eq.~(\ref{eq:beta-analyt}) in Appendi~\ref{sec:app-A}). Therefore, since  $\mathcal{F}\left(\infty\right)\rightarrow1$
at $H\gg\left|q_{\mathrm{p}}^{-1}\right|$, the reflected near-field signal generated
by the dipole $\vec{p}_{\omega}^{\prime}$ is the same as if there were no
graphene and only a bare wafer were present. Therefore, in the case of a wafer covered
by 2D material, $\hat{\beta}^{\prime}$-tensor is the same as for
the bare wafer, and it is given by Eq.~(\ref{eq:beta-prime-tensor}).
We can extend this conclusion to the case of 2D material on top of a wafer
with spatially non-homogeneous admittance. We have verified this for
the case of discontinuous doping junction (see Fig.~\ref{fig:s-snom-basics}b)
by matching calculated reflected near-field at $H\gg\left|q_{\mathrm{p}}^{-1}\right|,h$
and $\vec{E}$-field generated by the mirrored dipole $\vec{p}_{\omega}^{\prime}$
given by Eq.~(\ref{eq:beta-prime-tensor}). In the case of discontinuous
wafer junction geometry (see Fig.~\ref{fig:s-snom-basics}d), $\hat{\beta}^{\prime}$-tensor
becomes dependent on the $x$-coordinate, see Appendix~\ref{subsec:app-C-eps} for more details.

\section{Relationship between the plasmon reflection and the $S_2$ signal contrast\label{sec:sSNOM-reflection}}

To simulate s-SNOM response in the presence of graphene junctions, we employ a Fourier transform method for periodically repeated junctions, as shown in  Figs.~\ref{fig:periodic-geometry}(a-c). Technically, the Poisson equation solutions in these geometries are exactly the same as in the case of diffraction of the far-field incident on the sample~\cite{sanda1982gratings,semenenko2018broadening,klimenko2021broadening}. The only difference is that instead of the far-field, harmonics of the near-field generated by the tip's dipole are ``scattered'' on the sample (see Appendix\ref{sec:app-C} and Ref.~\cite{arsenin2010cylindric} for details). To eliminate spurious size effects, we choose the maximal height of the tip's dipole above the
sensed surface $h_{\mathrm{max}}$ to be much smaller than the system's
period $d=W+L$, and the plasmon's propagation lengths in the regions
between the junctions to be much smaller than the lengths
$W$ and $L$.

During the s-SNOM measurements, the tip senses the tip-launched standing wave, resulting from the interference between the tip and the edge-reflected plasmon. The interference pattern disappears at distances larger than the plasmon propagation length. As a result, the $S_{n}$ signal changes from a constant, which we call a relaxed signal intensity, to an oscillating signal with a growing amplitude as the tip approaches the junction. Examples of the simulated scans are
shown in Fig.~\ref{fig:S2-samples}a. This signal behavior is usually used in experiments to determine the coefficient of reflection from the plasmon junctions~\cite{wang2016soliton}. It is natural to assume that the envelope of the signal oscillations, normalized to its relaxed value, depends only on the junction's reflection coefficient.  The variation of the $S_2$ signal is considered to be proportional to the reflection coefficient, with the proportionality coefficient being independent of the junction type. This assumption enables one the calibration of the experimental setups using a junction with the known reflection.  For instance, it is well known that a graphene edge,
or a gap in graphene of the length bigger than about $\lambda_{\mathrm{p}}/4$,
almost entirely  reflects plasmons~\cite{nikitin2013foba,nikitin2014anomalous}.
We will show below that such assumptions are not always applicable.

In Fig.~\ref{fig:S2-samples}a, we demonstrate typical $S_{2}$ signals
for two junctions of graphene with discontinuous Fermi energy supported
by a homogeneous substrate. At the junctions, on the left side, the Fermi
energies are the same, while on the right side, the Fermi energies are chosen
to give the same plasmon reflection coefficients~\cite{khavasi2015diff-cond,semenenko2020reflection}. One can see that in this particular case, the described above assumptions are perfectly justified, i.e. the same reflection gives the same envelope. One of the most straightforward and reasonable estimation of the $S_2$ signal variation or contrast would be based on fitting the results to an expression like $\pm ae^{b\left(x-x_{0}\right)}$,
where $x_{0}$ is the junction position.  After finding the best-fit parameters $a$ and $b$, we could agree to
define the envelope's half-width as the fitted expression value at
some fixed position $x_{1}<x_{0}$ (e.g., at half of the plasmon wavelength to the junction). Our results show that the stated fitting procedure does not work well for processing  a broad collection of scans we have simulated. A much more reliable and accurate approach is to define the $S_2$ contrast as the absolute difference between the brightest (or the darkest) line and the relaxed signal. The choice depends on the relationship between the relaxed signals on the left and the right sides of the junction:
\begin{equation}
\varDelta S_{2}=\left\{ \begin{aligned}S_{2}^{\mathrm{BL}}-S_{2}^{\mathrm{Rx}},\;& S_{2}^{\mathrm{Rx}}>S_{2}^{\mathrm{Rx},2}\\
S_{2}^{\mathrm{Rx}}-S_{2}^{\mathrm{DL}},\;& S_{2}^{\mathrm{Rx}}<S_{2}^{\mathrm{Rx},2}
\end{aligned}
\right.,\label{eq:abs-dS2-def}
\end{equation}
where $S_{2}^{\mathrm{Rx}}$, $S_{2}^{\mathrm{Rx},2}$ are the $S_{2}$
signals on both sides of the junction measured at the distances
where the oscillations are relaxed, and $S_{2}^{\mathrm{BL}}$, $S_{2}^{\mathrm{DL}}$
are the brightest and the darkest lines of the signal scan on the
left side of the junction (see Fig.~\ref{fig:S2-samples}a). For the case of a gap in graphene junction, $S_{2}^{\mathrm{Rx},2}$ is taken as the absolute value of the $S_{2}$ signal from the bare wafer, i.e. in the absence of the junction. One should note that Eq.~\ref{eq:abs-dS2-def} is not the result of some self-consistent mathematical calculation. Instead we choose Eq.~\ref{eq:abs-dS2-def} definitions because of the similar constructions of $S_{2}^{\mathrm{BL}}$ and $S_{2}^{\mathrm{Rx}}$ have already been used in the experiments. We define $S_2$ contrast depending on the relation between the relaxed signals at both sides of a junction.  To show the conventionality of Eq.~\ref{eq:abs-dS2-def}, we developed a classification of the brightest and darkest lines on the left side of a junction as illustrated in Fig.~\ref{fig:S2-samples}b (see details in the caption).

In Figs.~\ref{fig:S2-refl-scaling}(a-d), we plot $\varDelta S_{2}$
normalized to $S_{2}^{\mathrm{Rx}}$ versus the absolute values of the reflection
from a junction for different types of junctions and various dielectric permittivities of the wafers supporting the junctions' left sides. Note that plasmon reflections can be calculated independently without involving an s-SNOM tip~\cite{khavasi2015diff-cond,bludov2018abrupt,semenenko2020reflection}.

It can be concluded from Figs.~\ref{fig:S2-refl-scaling}(a-d), that the assumption about the proportionality of the $S_2$ signal contrast with the reflection coefficient works reasonably well for $\varepsilon_{1}=1$ and $\varepsilon_{1}=3.9$. However, for the larger values of $\varepsilon_{1}$, multiple families of curves appear using all alternatives Eq.~\ref{eq:abs-dS2-def}. Moreover, even a single branch does not follow a straight line (see the filled markers). The deviations in Figs~\ref{fig:S2-refl-scaling}c,d are most prominent when the plasmon wavelength on the left side of a junction is smaller than on the right side (see star-markers). We relate this to the electric field generated by the long-plasmon-wavelength edge forming the junction. The short-plasmon-wavelength edge does not affect its counterpart because the edge's field penetration length into another side is about half of the plasmon wavelength on the edge's side~\cite{semenenko2020reflection}. Therefore, we find that the assumption about the proportionality between the signal contrast and the plasmon reflection breaks down for the case of a high-k dielectric substrate on the left side of the junction.

\section{Conclusion}

We applied a model for simulating the s-SNOM signal of 2D conducting for three types of 1D junctions. The model considers the tip oscillation, modulation, and demodulation of the near-field signal. This approach covers a broad range of experimentally relevant parameters. Our numerical approach is time efficient and allows us to explore a wide range of parameters to optimize plasmonic circuits. We provide an analytical model of s-SNOM of a conducting 2D material supported by a dielectric wafer that can be used for choosing a proper regime of the tip’s oscillation.

We have analyzed many simulated scans from different types of junctions and found that shapes of s-SNOM signal depend on multiple parameters, such that no simple fitting expression can cover all cases.  We have found that the assumption about the proportionality of the $S_2$ signal contrast and the reflection coefficient works well if a plasmonic junction is deposited on a low-k SiO$_2$ wafer, but it fails in the cases of high-k substrates. In addition, the assumption breaks down in most cases when the plasmon wavelength at the side of the junction, where the contrast is measured, is smaller than at the other side.

Our findings demonstrate that within the particular parameter space, the s-SNOM response can be used to extract material’s properties at the side of the junction, which is non-accessible for the s-SNOM tip. Our computationally efficient approach can generate a large amount of the training data sets for machine learning approaches for processing s-SNOM scans.

\section*{Acknowledgments}

We acknowledge support from the Vice President for Research and Economic Development (VPRED), SUNY Research Seed Grant Program, and the Center for Computational Research at the University at Buffalo~\cite{UBCCR}.

\appendix

\section{The case of homogeneous surface covered conducting 2D material}\label{sec:app-A}

In this section we present an analytical solution for the tensor's $\hat{\beta}$
components for the homogeneous infinitely thick dielectric wafer. This result is important for understanding the limitations of s-SNOM of 2D conducting materials and analyzing the s-SNOM scans of plasmonic junctions.

The Coulomb field produced by the linear dipole of the
tip on the sensing surface are given by:

\begin{equation}
E_{x}\left(x\right)=\frac{-4phx}{\left(x^{2}+h^{2}\right)^{2}},\quad E_{z}\left(x\right)=\frac{2p\left(h^{2}-x^{2}\right)}{\left(x^{2}+h^{2}\right)^{2}},
\end{equation}
where $h$ is the height of the dipole above the surface. The Fourier
transforms are: $e_{xq}=-2\pi i\cdot pqe^{-\left|q\right|h}$, $e_{zq}=2\pi\cdot p\left|q\right|e^{-\left|q\right|h}$.

The solution of the Poisson equation can be found as:


\begin{equation}
\varphi^{(r)}=\int\limits _{-\infty}^{\infty}\varphi_{q}^{(r,0)}e^{i\omega t-iqx-\left|q\right|z}\frac{dq}{2\pi},\quad\varphi^{(t)}=\int\limits _{-\infty}^{\infty}\varphi_{q}^{(t,0)}e^{i\omega t-iqx+\left|q\right|z}\frac{dq}{2\pi},
\end{equation}
where $\varphi^{(r)}$ is the potential distribution above the surface
(excluding the tip's dipole compound), and $\varphi^{(t)}$ is below
it. The boundary conditions at the surface lead to the following relations:

\begin{equation}
\begin{aligned} & -p\cdot2\pi iqe^{-\left|q\right|h}+iq\varphi_{q}^{(r,0)}=iq\varphi_{q}^{(t,0)}\\
 & \varepsilon_{0}\left[p\cdot2\pi\left|q\right|e^{-\left|q\right|h}+\left|q\right|\varphi_{q}^{(r,0)}\right]+\varepsilon_{1}\left|q\right|\varphi_{q}^{(t,0)}=4\pi\sigma_{\omega,q}
\end{aligned}
,\label{eq:bound-conds}
\end{equation}
where $\sigma_{\omega,q}$ are the Fourier components of the charge
density $\sigma\left(x,t\right)$ induced on the surface: $\sigma(x,t)=\int\limits _{-\infty}^{\infty}\sigma_{\omega,q}e^{i\omega t-iqx}\frac{dq}{2\pi}.$

We use the continuity equation $\frac{\partial\sigma}{\partial t}+\frac{\partial j}{\partial x}=0,$
where $j\left(x,t\right)$ is the linear current in the surface given
as $j_{\omega,q}=\gamma_{\omega}E_{x,\omega,q}$, where $\gamma_{\omega}$
is the surface conductivity, $E_{x,\omega,q}=iq\varphi_{q}^{(t,0)}$
is the E-field component producing the current in the surface, to obtain $\varphi_{q}^{(r,0)}$ and $\beta$ according to:

\begin{equation}
\beta=\int\limits _{-\infty}^{\infty}\frac{\varepsilon_{1}-\varepsilon_{0}-4\pi\left|q\right|i\gamma_{\omega}/\omega}{\varepsilon_{1}+\varepsilon_{0}-4\pi\left|q\right|i\gamma_{\omega}/\omega}\left|q\right|e^{-2h\left|q\right|}dq\label{eq:beta-gen}
\end{equation}

\noindent After some transformations, Eq.~(\ref{eq:beta-gen}) is reduced
to

\noindent
\begin{equation}
\begin{aligned} & \beta=\frac{1}{2\varepsilon_{0}h^{2}}\left[1-\frac{2\varepsilon_{0}}{\varepsilon_{0}+\varepsilon_{1}}\mathcal{F}\left(2q_{\mathrm{p}}h\right)\right],\;q_{\mathrm{p}}=\frac{\left(\varepsilon_{0}+\varepsilon_{1}\right)\omega}{4\pi i\gamma_{\omega}},\\
 & \mathcal{F}\left(\zeta\right)=\zeta^{2}e^{-\zeta}\left[\mathrm{Ei}\left(\zeta\right)+\pi i\right]-\zeta,\;\mathrm{Ei}\left(\zeta\right)=\int\limits _{-\infty}^{\zeta}\frac{e^{u}}{u}du,
\end{aligned}
\label{eq:beta-main}
\end{equation}

\noindent that for $\gamma_{\omega}=0$ gives the result similar to~\cite{hillenbrand2001basic}:~$\beta=\frac{\varepsilon_{1}-\varepsilon_{0}}{\varepsilon_{0}\left(\varepsilon_{1}+\varepsilon_{0}\right)}\frac{1}{2h^{2}}.$
Our analysis shows that in the case of the homogeneous conducting surface,
$\hat{\beta}$ is also proportional to $I$ as in the case of the bare
wafer, but now coefficient $\beta$ is different and depends on $z$
according to Eq.~(\ref{eq:beta-main}).

\begin{figure*}
\centering{\includegraphics[width=0.99\textwidth]{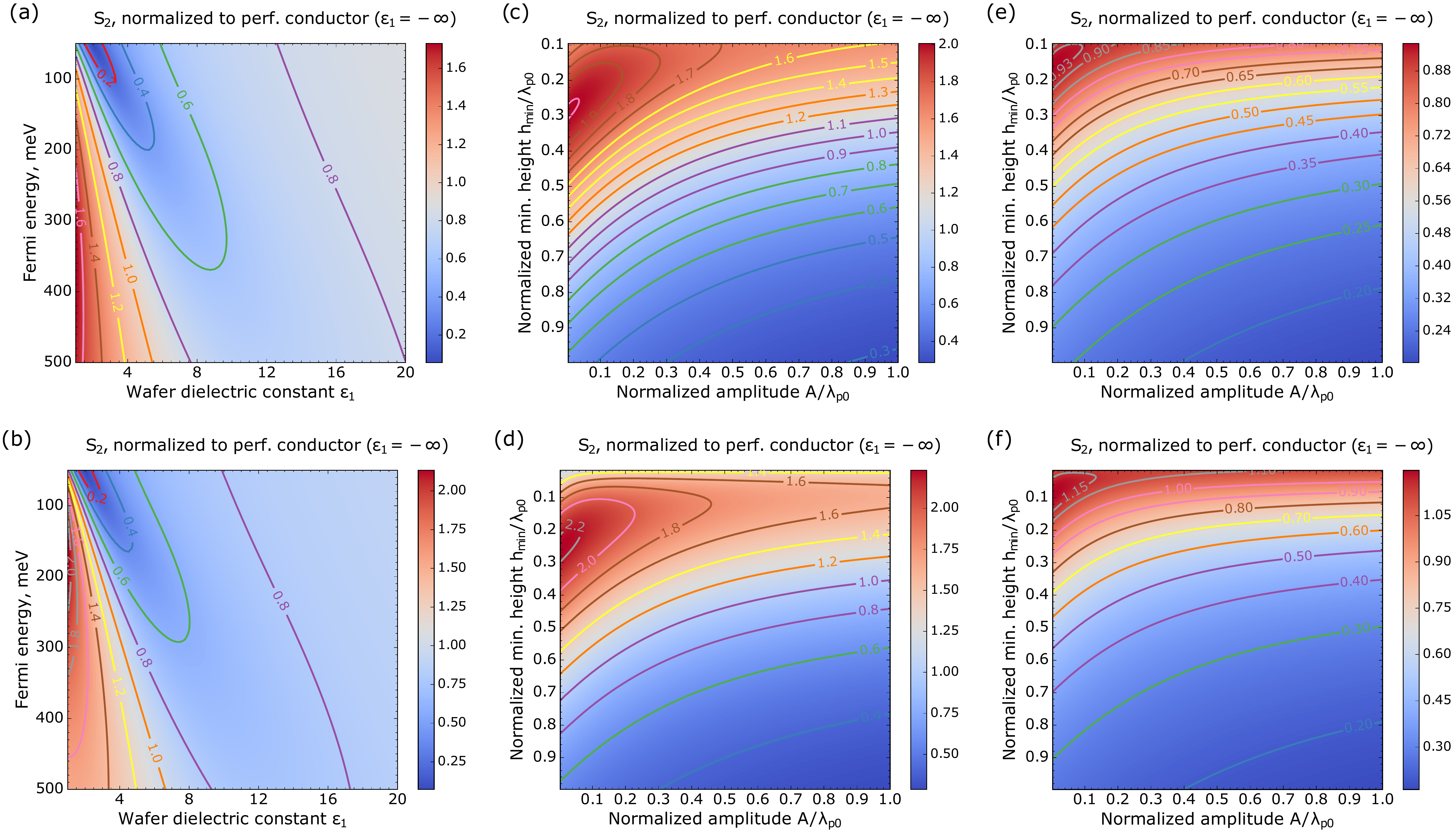}}

\caption{{\small{}(a) The simulated absolute value of the $S_{2}$ signal at frequency
$\omega=120$ meV from graphene with $\nu=10$ meV, calculated at
fixed $h_{\mathrm{min}}=35$ nm and $h_{\mathrm{max}}=80$ nm, and
plotted versus graphene Fermi energy and wafer's dielectric permittivity.
(c) Simulated absolute value of the $S_{2}$ signal at $\omega$ and $\nu$
as in (b), but for fixed Fermi energy of graphene $\mathcal{E}_{\mathrm{F}}=300$
meV and dielectric permittivity $\varepsilon_{1}=1.0$ below it, plotted
versus tip's dipole average height $h_{0}=0.5\left(h_{\mathrm{max}}+h_{\mathrm{min}}\right)$
and amplitude of mechanical oscillations $A=0.5\left(h_{\mathrm{max}}-h_{\mathrm{min}}\right)$
normalized to $\lambda_{\mathrm{p}0}=374$ nm. (e) The same as in
(c) but for $\nu=120$ meV. Tip's radius in panels (a), (c), and (e)
is $a=30$ nm. (b), (d), and (f) The same as in panels (a), (c) and
(e), correspondingly, but for tip radius $a=5$ nm. $\left|S_{2}\right|$
signals at every point of the maps at panels (a-f) are normalized
to the values of $\left|S_{2}\right|$ signal of the perfect conductor
at the corresponding $h_{0}$, $A$ and $a$.}}

\label{fig:s-SNOM-homogen}
\end{figure*}

Figs.~\ref{fig:s-SNOM-homogen} show the results of the absolute value
of the $S_{2}$ signal from graphene simulation normalized for the one
from the perfect conductor for the same tip's radius and CO$_{2}$
laser providing the incident beam with wavelength $10.6$ um that
corresponds to $120$ meV photon energy. Figs.~\ref{fig:s-SNOM-homogen}a,
b are plotted at standard fixed tip's oscillation regime ($5$ nm $\lesssim h-a\lesssim50$
nm) versus graphene Fermi energy $\mathcal{E}_{\mathrm{F}}$ and wafer's
dielectric permittivity $\varepsilon_{1}$. It is seen that the signal is non-monotonic and the tip’s radius does not affect the $S_2$ signal much. Collating Figs~\ref{fig:s-SNOM-homogen}c, d, one can
conclude that at higher ($\gtrsim10$) $Q$ the maximum of the $S_{2}$
signal is achieved approximately when the average height of the oscillating
dipole $h_{0}$ (see description to Fig.~\ref{fig:s-SNOM-homogen})
corresponds to the maximum of the real part of $\mathcal{F}\left(\zeta\right)$,
i.e. at $h_{0}/\lambda_{\mathrm{p}0}\approx0.3$. With decreasing
of $Q$, this ratio decreases (see the panels (e) and (f)) down to
zero (the case of $Q=0$ is not shown for not overloading the figure).
Note that despite $p_{\omega}$ dependence not only on $h/\lambda_{\mathrm{p}0}$,
$\varepsilon_{0}$ and $\varepsilon_{1}$ (see Eq.~(\ref{eq:beta-main}) and Eqs.~\eqref{eq:beta-tensor-def},
\eqref{eq:p-E-ref-rel-tensor}), but, at least,
on $a/h$, the scaling of $\left|S_{2}\right|$ versus $h_{0}/\lambda_{\mathrm{p}0}$
and $A/\lambda_{\mathrm{p}0}$ shown in Figs.~\ref{fig:s-SNOM-homogen}c-f
at the region where both arguments $\gtrsim0.4$ looks universal.
This happens because of the normalization to the $\left|S_{2}\right|$
from the perfect conductor simulated at the same tip's radius $a$,
optical frequency $\omega$, and oscillation regime ($h_{\mathrm{min}}$
and $h_{\mathrm{max}}$).

\section{s-SNOM of plasmonic junctions}\label{sec:app-B}

Simulation of the s-SNOM of a plasmonic junction can be done using
the same Eqs.~\eqref{eq:Sn-def} and \eqref{eq:beta-prime-tensor} in the main text
 in the case of sensing  a homogeneous surface. The only difference
now that the junction breaks homogeneity along the scanning direction
and $\hat{\beta}$ (see the definition at Eq.~\eqref{eq:beta-tensor-def} in the main text)
now depends on $x$-coordinate. Also, exact simulation of the signal,
even in the framework of our simplified model of dipole line tip and
reference wave considered to be parallel to the normal of the sensed
surface, must take into account $x$-component of the dipole moment
$\vec{p}_{\omega}$ induced in the tip. Thus, in the general case, in these kinds of simulations, the tensor value $\hat{\beta}$ cannot be reduced to just its single element $\beta=\beta_{zz}$ as it was
done in former s-SNOM models. However, in most, but not all cases,
which depend on the height of penetration of the evanescent  $E$-field
above the surface and the regime of tip's oscillation, one can neglect
all but $zz$-components of $\hat{\beta}$-tensor and use only $\beta_{zz}$
component instead of full Eq.~\eqref{eq:beta-prime-tensor}. Later
it will be shown when all the components of $\hat{\beta}$ must be
taken into account.

Our work presents  the numerical model of s-SNOM of three different
graphene plasmonic junctions: graphene doping discontinuity, a gap in
graphene, and discontinuity of dielectric permittivity of a wafer
supporting homogeneously doped graphene (see Figs.~1b-d in the main text).
As it was mentioned in the main text, for the
most part, s-SNOM signal simulation is reduced to the solution of
the forced problem, in terms of ordinary differential equations
theory, giving the relation between the dipole moment oscillations
with a fixed amplitude $\vec{p}_{\omega}$ and the $E$-field in the
system generated by it. The forced problem is solved in the periodic domain
using Fourier expansion of the unknown function as it's done in the
plain wave scattering on graphene grating~\cite{semenenko2018broadening}.
The only difference between  s-SNOM simulation and plain
wave scattering is in the boundary conditions for $E$-vector and normal $D$-vector components. Thus, if the $E$-field vector, generated
by the single tip located at $x=0$ with the dipole moment per
unit length $\vec{p}_{\omega}$, in the vicinity of the
sensed surface is $\vec{E}_{\mathrm{st}}\left(x,\vec{p}_{\omega}\right)$, then
the total electric field is:

\begin{equation}
\vec{E}_{0}\left(x,\vec{p}_{\omega}\right)=\sum\limits _{m=-M}^{M}\vec{E}_{\mathrm{st}}\left(x-md,\vec{p}_{\omega}\right)e^{-im\Phi},\label{eq:E0x-boundary}
\end{equation}
where $M\rightarrow\infty$ and $\Phi$ is the phase difference of
the dipole moment oscillations between adjacent tips. Once the unit
cell length $d$ is big enough as stated above, the result of the $S_{n}$
simulation does not depend on $\Phi$ and it is only needed to be set
non-zero to avoid singularity. For calculating the electric field compound generated by the tip’s dipole above the surface, one can use the formula analogous to the one for a point dipole, but for the case of the dipole line.  For the line of dipoles $\vec{p}_{\omega}$  along the $y$ direction and the position $\vec{h}$ the electric field at  $\vec{r}$ is given by:

\begin{equation}
\vec{E}_{\mathrm{st}}\left(\vec{r},\vec{p}_{\omega}\right)=\frac{4\left(\vec{p}_{\omega},\vec{\rho}\right)}{\left|\vec{\rho}\right|^{4}}\vec{\rho}-\frac{2\vec{p}_{\omega}}{\left|\vec{\rho}\right|^{2}},\label{eq:E-st-expr}
\end{equation}
where $\vec{\rho}=\vec{r}-\vec{h}$ and $\left(\vec{x},\vec{y}\right)$
is the scalar product of the vectors. As we transformed the original geometries
to the periodic ones, all the field distributions can be represented
in terms of the corresponding Fourier series. Thus, all the equations describing
the system are reduced to a set of the linear algebraic equations. The
tangential component of the $E$-vector and the normal component of $D$-vector
spatial distributions at the sensed surface can be represented as:

\begin{equation}
E_{0x}=\frac{1}{2}\sum\limits _{j=-N}^{N}\varepsilon_{0,j}e^{-iq_{j}x},\;D_{0z}=\frac{1}{2}\sum\limits _{j=-N}^{N}\varDelta_{0,j}e^{-iq_{j}x},\label{eq:E0xD0z-series}
\end{equation}
where $q_{j}=K+B_{0}j$, $K=\Phi/d$ is the quasi-momentum, $B_{0}=2\pi/d$
is the reciprocal unit cell period, and $\varepsilon_{0,j}$, $\varDelta_{0,j}$
are complex amplitudes of the Fourier harmonics calculated as:
\begin{equation}
\begin{aligned} & \varepsilon_{0,j}\left(\vec{p}_{\omega}\right)=\frac{2}{d}\int\limits _{-d/2}^{d/2}E_{0x}\left(x,\vec{p}_{\omega}\right)e^{iq_{j}x}\,dx,\\
 & \varDelta_{0,j}\left(\vec{p}_{\omega}\right)=\frac{2}{d}\int\limits _{-d/2}^{d/2}\varepsilon_{0}E_{0z}\left(x,\vec{p}_{\omega}\right)e^{iq_{j}x}\,dx.
\end{aligned}
\label{eq:eps0del0-series}
\end{equation}

Due to the linearity of the electromagnetic field equations with the dipole
moment, its $x$- and $z$-components can be considered separately.
We solve the self-consistent field excitation problem twice: for $\text{\ensuremath{\vec{p}_{\omega}=p_{x}\vec{i}}}$
and $\text{\ensuremath{\vec{p}_{\omega}=p_{z}\vec{k}}}$, where $\vec{i}$
and $\vec{k}$ are the unit vectors collinear to the $0X$ and $0Z$
directions, correspondingly. Thus, for the induced electric field
back-acting on the tip (see the main text accompanying Eq.~\eqref{eq:beta-tensor-def})
at $\ensuremath{\vec{p}_{\omega}=p_{x}\vec{i}}$ we obtain $\vec{E}_{\mathrm{ind}}^{(p_{x})}=p_{x}\left[e_{xx},\,e_{zx}\right]$
and at $\ensuremath{\vec{p}_{\omega}=p_{z}\vec{k}}$ it is $\vec{E}_{\mathrm{ind}}^{(p_{z})}=p_{z}\left[e_{xz},\,e_{zz}\right]$
where $e_{\alpha\beta}$ are components of vectors in $XZ$-plane
to be defined. Once the $e_{\alpha\beta}$ components are known, due
to the linearity and superposition principle, one can assemble from
them the $\hat{\beta}$ tensor:

\begin{equation}
\hat{\beta}=\begin{pmatrix}e_{xx} & e_{xz}\\
e_{zx} & e_{zz}
\end{pmatrix}.\label{eq:beta-tensor-components}
\end{equation}

Recently plasmon scattering problem for the three types of junctions was solved for the reflection coefficients~\cite{semenenko2020reflection}, which are used for the x-axises in Fig.~3d-f (in the main text).  
The components of $\hat{\beta}$ tensor in Eq.~(\ref{eq:beta-tensor-components})
are obtained using the same equations as in Ref.~\cite{semenenko2020reflection}
but in the presence of an external field which is unambiguously defined
by the Fourier components $\varepsilon_{0,j}$ and $\varDelta_{0,j}$. In other words, the forced excitation problem is being solved instead of
the eigenmodes calculation. The technical details of getting the solution
for the electromagnetic field (and plasmons in particular) are presented
in the next section.

\section{Solution of the forced problem of the electromagnetic field excitation}\label{sec:app-C}

The representation of electric field above the sensed surface is common for all three periodic geometries:

\begin{align}
\vec{E}^{(0)}\left(x,z\right) & =\vec{E}^{(0,\mathrm{ext})}\left(x,z\right)+\vec{E}^{(0,\mathrm{refl})}\left(x,z\right),\label{eq:E0-full}\\
\vec{E}^{(0,\mathrm{ext})}\left(x,z\right) & =\sum\limits _{n=-\infty}^{\infty}e^{-in\Phi}\vec{E}_{\mathrm{st}}\left(\vec{r}-n\vec{d},\vec{p}_{\omega}\right)\label{eq:E0-ext}\\
\vec{E}^{(0,\mathrm{refl})}\left(x,z\right) & =\sum\limits _{j=-\infty}^{\infty}a_{j}\begin{bmatrix}i\kappa_{0,j}\\
0\\
q_{j}
\end{bmatrix}e^{-iq_{j}x-\kappa_{0,j}z},\label{eq:E0-refl}
\end{align}
where $\mathrm{E}^{(0,\mathrm{ext})}$ is the electric field's component
produced by the array of tips (here $\vec{E}_{\mathrm{st}}$ is the
field from a single tip given by Eq.~(\ref{eq:E-st-expr}), $\Phi$
is the phase shift between adjacent unit cells, and $\vec{d}$ is
translation vector along $X$-axis of length $d=W+L$), and $\vec{E}^{(0,\mathrm{refl})}\left(x,z\right)$
is the total reflected near field that includes both the component
caused by polarization charges in the wafer supporting graphene, and
the one caused by the charges induced in graphene, which also can
be divided into quasistatic screening charges and oscillating charge
density of a plasmon. In the recent equation, $a_{j}$ are the amplitudes
of the reflected field Fourier components, $q_{j}=K+B_{0}j$ and $\kappa_{m,j}=\sqrt{q_{j}^{2}-\varepsilon_{m}\omega^{2}/c^{2}}$,
$\arg\kappa_{m,j}\in\left(-\pi/2,\,..\,\pi/2\right],$where $K=\Phi/d,$
$B_{0}=2\pi/d$ and $\varepsilon_{m}$ is the dielectric permittivity
of the medium labeled by index $m$.

In the cases of discontinuous doping and wafer geometries (Figs~3a,c), the solution of the problem
is found by matching  the boundary conditions for the electric fields
above and below the sensed surface, that is arranged at $z=0$. For
this purpose, from $\vec{E}^{(0,\mathrm{ext})}\left(x,z\right)$ 2D
dependence, we need to know only its slice at $z=0$, namely the tangential
component of $E$-vector $E_{0x}=\left.E_{x}^{(0,\mathrm{ext})}\right|_{z=0}$
and the normal component of $D$-vector $D_{0z}=\left.\varepsilon_{0}E_{z}^{(0,\mathrm{ext})}\right|_{z=0}$
spatial distributions. As the problem is solved in the spatial Fourier
domain, we expand $E_{0x}$ and $D_{0z}$ in Fourier series, and obtain
the corresponding amplitudes of the Fourier harmonics $\varepsilon_{0,j}$
and $\varDelta_{0,j}$ using Eqs~(\ref{eq:eps0del0-series}). In
the case of a gap-in-graphene junction, from the very beginning, we
solved the problem of excitation of quasistatic plasmons, i.e. found
the solution of Poisson equation instead of Maxwell equations, using
the Coulomb law directly. For this reason, only $E_{0x}$ component
and its Fourier series $\varepsilon_{0,j}$ was required from the
whole electric field vector distribution $\left.\vec{E}^{(0,\mathrm{ext})}\right|_{z=0}$.

The resulting solution for plasmons in graphene is found in terms of Fourier series
$u_{j}$ of the spatial distribution of complex amplitudes $j_{\omega}\left(x\right)$
of the linear current oscillations:

\begin{equation}
j\left(x,t\right)=\mathrm{Re}\,e^{i\omega t}j_{\omega}\left(x\right),\quad j_{\omega}=\frac{1}{2}\sum\limits _{j=-N}^{N}u_{j}e^{-iq_{j}x}.
\end{equation}

In the following we show the resulting equation systems to be solved for the cases of all three types of junctions.

\subsection{Discontinuous doping}\label{subsec:app-C-muc}

\begin{equation}
\begin{aligned} & \sum\limits _{m=-N}^{N}\left[\delta_{jm}-\frac{4\pi i\gamma_{\omega}}{\omega}\gamma_{jm}\kappa_{0,m}R_{m}\right]u_{m}=\\
 & =\gamma_{\omega}\sum\limits _{m=-\infty}^{\infty}\gamma_{jm}\left[i\kappa_{0,m}F_{m}+\varepsilon_{0,m}\right],
\end{aligned}
\end{equation}
where
\begin{equation}
F_{j}=\left[\varepsilon_{1}\frac{q_{j}\Xi_{j}^{\prime}\varepsilon_{0,j}}{B_{0}\gamma_{1,j}}-\varDelta_{0,j}\right]\left/\left[\left(\varepsilon_{0}-\varepsilon_{1}\frac{i\kappa_{0,j}\Xi_{j}^{\prime}}{B_{0}\gamma_{1,j}}\right)q_{j}\right]\right.
\end{equation}
for $j\neq0$, otherwise:
\begin{equation}
F_{0}=\varepsilon_{1}\frac{\Xi_{j}^{\prime}\varepsilon_{0,0}}{B_{0}\gamma_{1,0}}\left/\left[\varepsilon_{0}-\varepsilon_{1}\frac{i\kappa_{0,0}\Xi_{0}^{\prime}}{B_{0}\gamma_{1,0}}\right]\right.,
\end{equation}
and the values $\gamma_{jm}$, $R_{m}$, $\Xi_{j}^{\prime}$ and $\gamma_{1,j}$
are described in the Appendix of the Ref.~\cite{semenenko2020reflection}.
After $u_{j}$ harmonics are found, the Fourier harmonics $a_{j}$ of
the total reflected near field $\vec{E}^{(0,\mathrm{refl})}$ can
be calculated using the following equation:
\begin{equation}
a_{j}=\frac{4\pi}{\omega}R_{j}u_{j}+F_{j}.
\end{equation}

\subsection{Discontinuous wafer}\label{subsec:app-C-eps}
Here we use a similar approach to Ref.~\cite{sanda1982gratings}:
\begin{align}
 & \begin{aligned} & \sum_{\varkappa}\left(\varepsilon_{0}\frac{q_{j}}{\kappa_{0,j}}\varepsilon_{\varkappa j}-4\pi q_{j}\frac{i\gamma_{\omega}}{\omega}\varepsilon_{\varkappa j}-i\varDelta_{\varkappa j}\right)C_{\varkappa}=\\
 & =\varepsilon_{0}\frac{q_{j}}{\kappa_{0,j}}\varepsilon_{0,j}-i\varDelta_{0,j},\;u_{j}=\gamma_{\omega}\sum_{\varkappa}\varepsilon_{\varkappa j}C_{\varkappa},
\end{aligned}
\label{eq:C-kappa}\\
 & \quad i\kappa_{0,j}a_{j}+\varepsilon_{0,j}=\sum_{\varkappa}\varepsilon_{\varkappa j}C_{\varkappa},\label{eq:a0j-eps}
\end{align}
where $\varkappa=\left\{ \varkappa_{l}\right\} $ is the set of roots
of the spectral equation described in the Ref.~\cite{semenenko2020reflection},
and the summations over $\varkappa$ mean the summation over the
root's index $l$ which is omitted in the expressions after $\Sigma$
for better readability; values $\varepsilon_{\varkappa j}$, $\varDelta_{\varkappa j}$
and $C_{\varkappa}$ are also described in the Appendix of the Ref.~\cite{semenenko2020reflection}.
After $C_{\varkappa}$ coefficients are found, the Fourier harmonics $a_{j}$
of the total reflected near field $\vec{E}^{(0,\mathrm{refl})}$ can
be calculated using Eq.~(\ref{eq:a0j-eps}). To calculate $\hat{\beta}^{\prime}$-tensor
from Eq.~\eqref{eq:beta-analyt} in the main text, we solve Eqs.~(\ref{eq:C-kappa})
for horizontally and vertically oriented exciting dipole moment $\vec{p}_{\omega}=1\cdot\vec{i}$
and $\vec{p}_{\omega}=1\cdot\vec{k}$, then knowing $a_{j}$ we calculate
$x$- and $z$-components of electric field $E_{x,j}^{(p_{x})}$,
$E_{z,j}^{(p_{x})}$ and $E_{x,j}^{(p_{z})}$, $E_{z,j}^{(p_{z})}$
using Eq.~(\ref{eq:E0-refl}) at some certain height $\sim2\div3z$
above the sample and at $x=0$. Then, we calculate the base harmonics
of $\vec{E}$-field produced by the same $\vec{p}_{\omega}$-dipole
located at the same $x$-coordinate and negative $z$-coordinate:

\begin{equation}
\begin{aligned} & E_{x0,j}^{(p_{x})}=\varepsilon_{0,j}\left(1\cdot\vec{i}\right)e^{-\kappa_{0,j}z},\\
 & E_{z0,j}^{(p_{x})}=-\varDelta_{0,j}\left(1\cdot\vec{i}\right)e^{-\kappa_{0,j}z}/\varepsilon_{0},\\
 & E_{x0,j}^{(p_{z})}=-\varepsilon_{0,j}\left(1\cdot\vec{k}\right)e^{-\kappa_{0,j}z},\\
 & E_{z0,j}^{(p_{z})}=\varDelta_{0,j}\left(1\cdot\vec{k}\right)e^{-\kappa_{0,j}z}/\varepsilon_{0},
\end{aligned}
\label{eq:mirrored-dipole-base}
\end{equation}
where $\varepsilon_{0,j}\left(\vec{p}_{\omega}\right)$ and $\varDelta_{0,j}\left(\vec{p}_{\omega}\right)$
are given by Eqs.~(\ref{eq:eps0del0-series}). Then the components
of $\hat{\beta}^{\prime}$-tensor are found from the following equations:

\begin{equation}
\begin{pmatrix}E_{x0,j}^{(p_{x})} & E_{x0,j}^{(p_{z})}\\
E_{z0,j}^{(p_{x})} & E_{z0,j}^{(p_{z})}
\end{pmatrix}\begin{pmatrix}\beta_{xx}^{\prime} & \beta_{zx}^{\prime}\\
\beta_{xz}^{\prime} & \beta_{zz}^{\prime}
\end{pmatrix}=\begin{pmatrix}E_{x}^{(p_{x})} & E_{x}^{(p_{z})}\\
E_{z}^{(p_{x})} & E_{z}^{(p_{z})}
\end{pmatrix}.\label{eq:beta-prime-solver}
\end{equation}

As a result, $\hat{\beta}^{\prime}$-tensor is found with the high accuracy,
and its value remains constant while $z\ll d$.

\subsection{Gap in graphene}\label{subsec:app-C-gap}
The same approach for the case of similar and non-periodic geometry was considered before in Ref.~\cite{arsenin2010cylindric}.
\begin{align}
 & \sum\limits _{l=-N}^{N}\left(\delta_{jl}-\frac{i\gamma_{\omega}}{\omega}\frac{\pi B_{0}^{\prime}}{\varkappa}M_{jl}\right)u_{l}^{\prime}=\frac{1}{\varkappa}\gamma_{\omega}\varepsilon_{0,j}^{\prime},\\
 & \sum\limits _{l=-N}^{N}\left(-1\right)^{l}u_{l}^{\prime}=0,\;\varkappa=\frac{\varepsilon_{0}+\varepsilon_{1}}{2},\;B_{0}^{\prime}=\frac{2\pi}{W},
\end{align}
where
\begin{equation}
M_{jl}=M_{jl}^{(0)}+\sum\limits _{k=1}^{K}\left(M_{jl}^{(k,\mathrm{a})}\cos k\Phi-M_{jl}^{(k,\mathrm{s})}\sin k\Phi\right),
\end{equation}
where $\varepsilon_{0,j}^{\prime}$ and $u_{l}^{\prime}$ are Fourier
expansions of $E_{0x}(x)$ and $j_{\omega}\left(x\right)$ on shorter
range $\left(-W/2\,..\,W/2\right)$ than their actual period $d$:

\begin{equation}
\begin{aligned} & \varepsilon_{0,j}^{\prime}=\frac{2}{W}\int\limits _{-W/2}^{W/2}E_{0x}\left(x\right)e^{iB_{W}jx}\,dx,\\
 & u_{j}^{\prime}=\frac{2}{W}\int\limits _{-W/2}^{W/2}j_{\omega}\left(x\right)e^{iB_{W}jx}\,dx,
\end{aligned}
\end{equation}
$K$ is the number of neighbors of a graphene ribbon are taken into
account and the values $M_{jl}^{(0)}$, $M_{jl}^{(k,\mathrm{a})}$,
and $M_{jl}^{(k,\mathrm{s})}$ are described in the Appendix of the
Ref.~\cite{semenenko2020reflection}. After $u_{j}^{\prime}$
complex harmonics are calculated, they are transformed to the $s_{j}^{\prime}$
harmonics using the charge conservation equation $\frac{\partial\sigma}{\partial t}+\frac{\partial j}{\partial x}=0$,
or $s_{j}^{\prime}=jB_{0}^{\prime}/\omega\cdot u_{j}^{\prime}$, and
then $s_{j}^{\prime}$ are transformed to the harmonics of the Fourier
expansion of the periodic function $\sigma_{\omega}\left(x\right)$
in the Bloch representation of the charge density oscillation $\sigma\left(x,t\right)$
in the whole sensed surface:

\[
\sigma\left(x,t\right)=\mathrm{Re}\,e^{i\omega t}\sigma_{\omega}\left(x\right),\quad\sigma_{\omega}=\frac{1}{2}\sum\limits _{j=-N}^{N}s_{j}e^{-iq_{j}x},
\]
where $q_{j}=K+B_{0}j$ ($K$ and $B_{0}$ are defined above).

Once the surface charge density distribution $\sigma\left(x,t\right)$
and its Fourier expansion $s_{j}$ are known, one can solve the Poisson
equation and find the corresponding to it electrostatic potential
distribution in the regions above and below the surface $z=0$ in
the following form:

\begin{equation}
\varphi^{\mathrm{surf}}\left(x,z\right)=\frac{1}{2}\sum\limits _{j=-N}^{N}\varphi_{j}e^{-iq_{j}x\mp\left|q_{j}\right|z},
\end{equation}
where the sign ``-'' corresponds to the region $z\geq0$, and ``+''
to $z<0$. Solution of the Poisson equation gives:

\[
\varphi_{j}=\frac{2\pi}{\varkappa}\frac{s_{j}}{q_{j}}.
\]
Knowing $\varphi^{\mathrm{surf}}$, one can calculate the corresponding
to it electric field $\vec{E}^{\mathrm{surf}}=-\frac{\partial\,}{\partial\vec{r}}\varphi^{\mathrm{surf}}$,
and to obtain the total near electric field reflected from the sensed
sample, we need to add the component associated with the polarization
charges in the wafer supporting the graphene: $\vec{E}^{(0,\mathrm{refl})}=\vec{E}^{\mathrm{surf}}+\vec{E}^{\mathrm{bulk}}$,
where

\[
\vec{E}^{\mathrm{bulk}}=\frac{\varepsilon_{1}-\varepsilon_{0}}{\varepsilon_{1}+\varepsilon_{0}}\sum\limits _{j=-\infty}^{\infty}\begin{bmatrix}-\varepsilon_{0,j}\\
0\\
\frac{\varDelta_{0,j}}{\varepsilon_{0}}
\end{bmatrix}e^{-iq_{j}x-\left|q_{j}\right|z},
\]
where $\varepsilon_{0,j}$ and $\varDelta_{0,j}$ are calculated above using Eq.~(\ref{eq:eps0del0-series}).

\bibliographystyle{unsrt} 
\bibliography{references}

\end{document}